\def\kms{\relax \ifmmode {\,\rm km\,s}^{-1}\else \,km\,s$^{-1}$\fi}

\def\farcs{\hbox{$.\!\!^{\prime\prime}$}}
\def\secd#1.#2{ #1\farcs#2 }               % seconds over decimal point

\def\mincir{\ \raise-2.truept\hbox{\rlap{\hbox{$\sim$}}\raise5.truept
    \hbox{$<$}\ }}
\def\magcir{\ \raise-2.truept\hbox{\rlap{\hbox{$\sim$}}\raise5.truept
    \hbox{$>$}\ }}
\def\gr{$^\circ$}

\def\nii{[N~{\sc ii}]}
\def\teoiii{T$_{\rm e}$[O III]}
\def\tenii{T$_{\rm e}$[N II]}
\def\oiii{[O~{\sc iii}]}

\def\oii{[O~{\sc ii}]}

\def\hb{H$\beta$}

\def\hii{H~{\sc ii}}
\def\heii{He~{\sc ii}}

\def\hei{He~{\sc i}}
\def\sii{[S~{\sc ii}]}

\def\cbeta{c({H$\beta$})}
\def\ne{$n_{e}$}
\def\nii{[N~{\sc ii}]}

\def\oiii{[O~{\sc iii}]}

\def\oii{[O~{\sc ii}]}

\def\hb{H$\beta$}

\def\hii{H~{\sc ii}}
 \def\heii{He~{\sc ii}}

\def\hei{He~{\sc i}}
\def\sii{[S~{\sc ii}]}

\def\ne{$n_{e}$}

\documentclass[preprint2]{aastex}

\shorttitle{The PN metallicity gradient in M33}
\shortauthors{Magrini, Stanghellini, Villaver}

\begin{document}

\title{
  The planetary nebula population of M33 and its metallicity gradient:
  A look into the galaxy's distant past   
  \footnote{Observations reported here were obtained at the MMT 
  Observatory, a joint facility of the Smithsonian Institution 
  and the University of Arizona.}}

\author{Laura Magrini\altaffilmark{1,2}}
\affil{Istituto Nazionale di Astrofisica, Osservatorio Astrofisico di Arcetri}
\affil{Istituto Nazionale di Ottica Applicata,  Firenze, I 50125, Italy}
\email{laura@arcetri.astro.it}

\author{Letizia Stanghellini\altaffilmark{3}}
\affil{National Optical Astronomy Observatories, Tucson, AZ 85719}
\email{lstanghellini@noao.edu}

\and

\author{Eva Villaver\altaffilmark{4,5}} %EVA
\affil{Space Telescope Science Institute, Baltimore, MD 21218}
\affil{Affiliated with the Hubble Space Telescope Division
of the European Space Agency} %EVA
\email{villaver@stsci.edu}

\begin{abstract}
The Planetary Nebula (PN) population of M33 is studied via multi-fiber
spectroscopy with Hectospec at the MMT.  In this paper we present the
spectra of 102 PNe, whereas plasma diagnostic and chemical abundances
were performed on the 93 PNe where the necessary diagnostic
lines were measured.

About 20$\%$ of the PNe are compatible with being Type I; the rest of
the sample is the progeny of an old disk stellar population, with main
sequence masses M$<$3M${_\odot}$ and ages t$>$0.3~Gyr.

By studying the elemental abundances of the PNe in the M33 disk we
were able to infer that: (1) there is a tight correlation between O/H
and Ne/H, broadly excluding the evolution of oxygen; (2) the average
abundances of the $\alpha$-elements are consistent with those of \hii\
regions, indicating a negligible global enrichment in the disk of M33
from the epoch of the formation of the PN progenitors to the present
time; (3) the radial oxygen gradient across the M33 disk has a slope
of -0.031$\pm$0.013 dex kpc$^{-1}$, in agreement, within the errors,
with the corresponding gradient derived from \hii\ regions. Our
observations do not seem to imply that the metallicity gradient across
the M33 disk has flattened considerably with time.  We report also the
discovery of a PN with Wolf-Rayet features, PN039, belonging the class
of late [WC] stars

\end{abstract}


\section{Introduction}

The galaxy M33 (NGC~598) is one of the closest spiral galaxies of the
Local Group.  The measured distance to M33 ranges from 730~kpc
(Christian \& Schommer~1987) to 910~kpc (Kim et al.~2002), with the 
most recent estimates of Sarajedini et al.~\citep{sarajedini06} and
Bonanos et al.~\citep{bonanos06} in the mid-range.  The proximity of M33,
together with its large angular size (optical size 53'$\times$83',
Holmberg 1958), and its intermediate inclination ($i$=53\arcdeg),
allows detailed studies of its stellar populations and ionized
nebulae.

M33 is a galaxy rich in both PNe and \hii\ regions.  The population of
PNe in M33 was early investigated using an objective-prism
survey by Ford~\cite{ford83} and Lequeux et al.~\cite{lequeux87}. The advent
of wide-field CCD cameras allowed 
deeper surveys. Magrini et al.~\cite{magrini00}
identified 131 candidate PNe. More recently, 
Ciardullo et al.~\cite{ciardullo04} confirmed a large number of
previously detected PNe and identified new candidates, leading
to the current number of spectroscopically confirmed PNe: 138 in the M33 disk 
and 2 in its halo.

Due the closeness of M33 and the brightness of the giant \hii\ regions,
their spectroscopy was obtained since the 70s. Searle \cite{searle71}
presented spectrophotometry of eight \hii\ regions, and further
spectroscopic studies were carried on by Smith~\cite{smith75}, Kwitter
\& Aller~\cite{kwitter81}, and by V\'{\i}lchez et
al.~\cite{vilchez88}.  Several
catalogs of the M33 \hii\ regions have been published, such as those
by Courtes et
al.~\cite{courtes87}, Calzetti et al.~\cite{calzetti95}, Wyder et
al.~\cite{wyder97}, and Hodge et al.~\cite{hodge99}.  Recently, the Local
Group Census (LGC, Corradi \&
Magrini~2006) and the Local Group Survey (LGS, Massey et
al.~2007) 
have provided deep narrow- and broad-band $\sim$2 square degree coverage 
of M33, disclosing a conspicuous number of new emission-line objects located
at all galactocentric distances.   

Given the wealth of information on its PN and \hii\ region
populations, M33 is indeed an ideal candidate to test chemical
evolution predictions. The aim of the spectroscopic studies of PNe and
\hii\ regions is devoted  both to the individual studies of these
objects, and to draw the radial metallicity gradient across the
disk. In particular, PNe and \hii\ regions have two very
different formation ages, and the comparison of these two populations
provides insight on the 
evolution of the host galaxy. 
PN progenitors are low- and intermediate-mass stars (LIMS), with masses between 
1 and 8 M$_{\odot}$, which must have been formed between
$\sim$3$\times10^7$ yr and 10 Gyr ago (Maraston 2005). Instead \hii~regions
are very young.  

During their evolution, LIMS do not modify, at
least at 
zeroth approximation, the 
composition of $\alpha$- elements such as oxygen, neon, argon, and
sulfur. 
These elements are produced mainly from the nucleosynthesis of Type II
supernovae and maintain their capability to testify their original presence
in the interstellar cloud that gave birth to the PN progenitor.
While there are indications that for extremely low metallicities both oxygen and neon
could be modified, as it has been observed both in the SMC and in other dwarf galaxies (Leisy and
Dennefeld 2006; Magrini et al. 2005; Kniazev et al. 2008), 
at the metallicity of M33 one does not expect any nucleosynthesis activity
involving these elements (Marigo 2001).

On the other hand, the helium, nitrogen, and carbon abundances 
measured in PNe do not correspond to those at the time of the
progenitor's formation, since these elements are synthesized in LIMS.  
These elements hence give
information on LIMS evolution accordingly to
their initial mass and metallicity, and, at given metallicity, are
helpful to constrain the PN progenitor mass and age.

The M33 metallicity gradient derived from $\alpha$-element abundances has been
the focus of several studies already, 
both involving PNe and
\hii\ regions. The advantage of M33 with respect to our Galaxy, as 
a playground for gradients, is that PNe in M33 have 
well determined galactocentric distances (with relative errors within
5$\%$) compared to the large indetermination of Galactic PN distances
(Stanghellini et al.~2008). Magrini et
al.~\cite{magrini04} using PN spectroscopy in M33 determined the elemental
abundances of 11 
PNe, and Magrini et al.~(2007b) (hereafter M07b) derived an oxygen radial gradient of
$\Delta$(O/H) / $\Delta$R = -0.11$\pm$0.04~dex~kpc$^{-1}$, where R is the
galactocentric distance. The sample of 11 PNe was too small for 
a definite answer to the PN gradient of M33, especially given the paucity of
PNe observed at large galactocentric distances. 

The metallicity gradient of M33 using \hii\ regions has been obtained by many
authors, with broadly different results.  The first studies by
Smith~\cite{smith75}, Kwitter \& Aller~\cite{kwitter81}, and
V\'{\i}lchez et al.~\cite{vilchez88} agreed on a steep oxygen
gradient. Garnett et al.~\cite{garnett97} also obtained a steep
gradient, with $\Delta$(O/H) / $\Delta$R =
-0.11$\pm$0.02~dex~kpc$^{-1}$ by homogeneously compiling 
published data.  Recent determinations, such as those
by Crockett et al.~\cite{crockett06}, Magrini et
al.~\cite{magrini07a}, and Rosolowsky et al.~\cite{rs08} (hereafter
M07a and RS08, respectively) seem to converge to a much shallower
gradient, respectively deriving $\Delta$(O/H) / $\Delta$R =
-0.012$\pm$0.011, -0.054$\pm$0.011, and
-0.027$\pm$0.012~dex~kpc$^{-1}$.  The most recent result by Rubin et
al.~\cite{rubin08}, based on Ne/H and S/H, gives -0.058$\pm$0.014 and
-0.052$\pm$0.021~dex~kpc$^{-1}$, respectively.  The RS08 sample is the
largest to date, and should be used preferentially, since the results from small
samples might emphasize the scatter rather than the slope.

Metallicity gradients in M33 have also been estimated form young giant
stars (Herrero et al.~1994, McCarthy et al.~1995, Venn et al.~1998,
Monteverde et al.~1997, 2000, Urbaneja et al.~2005), and from AGB (Cioni et
al.~2008), RGB stars (Stephens et al.~2002, Kim et al~2002, Galletti
et al.~2004, Tiede et al.~2004, Brooks et al.~2004, Barker et
al.~2006), and Cepheids (Beaulieu et al.~2006).

%why gradient? 
One of the most discussed questions about the metallicity gradient in
disk galaxies is how it evolves with time.  Chemical evolution models
(see e.g. M07b for a review of M33 models) predict
different temporal behaviors of the metallicity gradient depending
on assumptions such as gas inflow and outflow rate, and star and cloud
formation efficiencies.  Observations are needed to constrain these
theoretical scenarios, but so far they have been insufficient,
especially for the old populations. Comparing different sets of
results for the young stellar populations, such as \hii\ regions,  and
the old population, such as AGB and RGB stars, is also delicate, since the techniques of observing and
analyzing nebulae and stars are very different, each with its own collection
of uncertainties.

The idea behind the observations leading to this paper is to study the
chemical and physical properties of a large number of PNe and \hii\
regions using the same set of observations, the same data reduction and
analysis techniques, and identical abundance determination methods,  in
order to avoid all biases due to the stellar vs. nebular analysis.
The chemical properties of PNe and \hii\ regions will provide us with snapshots at
two epochs in the life-time of M33, in particular of its metallicity
gradient. In
the present paper we show the results obtained from the PN data.
We have obtained
the first sizable sample of M33 PNe with uniformly derived abundances in
order to build the first sound gradient determination 
from PNe. In 
a forthcoming paper we will present our own \hii\ region results, and we will
compare them with the PN results presented here.

This paper is organized as follows: in $\S$~\ref{sect_obs} we present
the observations and data reduction, in $\S$~\ref{sect_abu} we
discuss the plasma diagnostics used and how we determine the abundances of
the PNe, and in $\S$~\ref{sect_pne} we describe the PN properties.
In  $\S$~\ref{sec_comp} we compare the results with the PN populations in other galaxies.   The chemical
abundance gradients and their implications in the evolution of M~33
are discussed in $\S$~\ref{sect_grad} and in $\S$~\ref{sect_discu}.
Finally in $\S$~\ref{sect_conclu} we summarize our results, and
present the conclusions of this work.

\section{Observations and data reduction}
\label{sect_obs}

We obtained spectra of 102 PNe and 48 \hii\ regions in M33 using the
MMT Hectospec fiber-fed spectrograph (Fabricant et al.~2005) which is equipped
with an Atmospheric Dispersion Corrector. The spectrograph was used
with a single setup: 270 mm$^{-1}$ grating at a dispersion of 1.2 \AA
~pixel$^{-1}$.  The resulting total spectral coverage ranged from
approximately 3600
\AA\ to 9100 \AA, thus including the basic emission-lines necessary
for the determination of their
physical and chemical properties.  The instrument deploys 300 fibers
over a 1\gr diameter field of view and the fiber diameter is $\sim$ 1.5\arcsec\  (6 pc using a distance of 840 kpc to M~33).  The 102 PNe were
selected from the catalog of Ciardullo et al.~\cite{ciardullo04},
including four new PNe discovered therein at large galactocentric radii.

In Table~\ref{tab_pn} we present the list of the observed PNe. Column (1) gives
the identification number from Ciardullo et
al.~\cite{ciardullo04} except for PNe 153 to 156 which are from LGC
observations (Corradi \& Magrini 2006);  
columns (2) and (3) give the equatorial coordinates, RA and DEC at J2000.0, and column (4) gives the
\oiii\ magnitude  from Ciardullo et al.~\cite{ciardullo04} 
following the definition by Jacoby~\cite{jacoby89}.

\begin{deluxetable}{cccc}
\tabletypesize{\scriptsize}
%\rotate
\tablecaption{Observed PNe}
\tablewidth{0pt}
\tablehead{
\colhead{Id} & \multicolumn{2}{c}{RA J2000.0 DEC} & \colhead{M [OIII]}\\
%                      &\multicolumn{2}{J2000.0} 
(1)                & (2)                 & (3)                             & (4) 
}
\startdata
PN001 &   1:32:09.04 & 30:22:05.700 & 23.51 \\  
PN002 &   1:32:26.54 & 30:25:49.800 & 23.53 \\  
PN003 &   1:32:38.03 & 30:24:00.603 & 21.82 \\  
\enddata
\label{tab_pn}
\tablecomments{(1) Identification number from Ciardullo et
al.~\cite{ciardullo04}, except for PNe 153 to 156, which are from LGC observations (Corradi \& Magrini 2006); 
(2), (3)  galactic coordinates at J2000.0;  (4)  \oiii\ magnitude 
following the definition by Jacoby~\cite{jacoby89}. Table~1 is published in its entirety in the 
electronic edition of the {\it Astrophysical Journal}.  A portion is 
shown here for guidance regarding its form and content. 
}
\end{deluxetable}

The \hii\ regions of our sample were chosen either based on their \oiii\
brightness, or among those whose chemical abundances had already been
measured by others so we can  use them as a control sample. In this paper we
only use the data of the \hii\ regions to 
test our procedures by comparing our fluxes with those in the literature.

We used a large number of fibers to take sky spectra for sky-subtraction.
In order to optimize the sky-subtraction we selected low diffuse-emission
areas on 
the face of M33 
from the INT H$\alpha$ and \oiii-continuum frames (Magrini et al. 2000).
Since the continuum from the PN central star
can not be detected at the M33 distance any continuum emission in the spectra
has to be arising from an underlying unresolved stellar population.
However, the continuum emission in our spectra is in most cases negligible, and
therefore the Balmer absorption lines coming from the stellar background are unimportant. 

Five 1800~s exposures were taken on
the night of October 13, 2007, and three additional 1800~s exposures
of the same field were obtained on November 12, 2007.  The airmass
during the observations ranged between 1.07 and 1.4 during the night
of October 13, and from 1.3 to 1.5 on November 12.  The seeing was 1.2
\arcsec\ and 1.7\arcsec, respectively.  Several dome-flat and sky-flat
exposures were obtained during the nights of observations to perform
the data reduction.  Arc exposures with the calibration lamp He-Ne-Ar
were taken for wavelength calibration.

The spectra were reduced using the Hectospec package. All observations
were bias subtracted, overscan corrected, and trimmed. The science
exposures were flat-fielded and combined together to eliminate cosmic
rays, and the one-dimensional spectra were extracted and wavelength
calibrated.  The relative flux calibration was done observing the
standard star Hiltm600 (Massey et al.~1988) during the nights of
October 15 and November 27.  The standard star was observed with
airmass $\sim$1.1.  The emission-line fluxes were measured with the
package SPLOT of IRAF\footnote{IRAF is distributed by the National
Optical Astronomy Observatory, which is operated by the Association of
Universities for Research in Astronomy (AURA) under cooperative
agreement with the National Science Foundation}.  Errors in the fluxes
were calculated taking into account the statistical error in the
measurement of the fluxes, as well as systematic errors of the flux
calibrations, background determination, and sky subtraction.

The observed line fluxes were corrected for the effect of the
interstellar extinction using the extinction law of
Mathis~\cite{mathis90} with $R_V$=3.1.  

We derived  \cbeta, the logarithmic
nebular extinction, by using the weighted average of the
observed-to-theoretical Balmer ratios of 
H$\alpha$, H$\gamma$, and H$\delta$ to H$\beta$  (Osterbrock \& Ferland 2006). 

Table 2 gives the results of our line measurements and extinction
corrections. Column (1) gives the PN name; column (2) gives the nebular
extinction coefficient
\cbeta\ with its error;  columns (3) and (4) indicate the emitting ion and
the rest-frame wavelength 
in \AA;  columns (5), (6), and (7) give the measured line fluxes
(F$_{\lambda}$), their absolute  errors ($\Delta$(F$_{\lambda}$),  and finally the extinction 
corrected fluxes (I$_{\lambda}$). Both F$_{\lambda}$
and I$_{\lambda}$ are normalized to \hb=100. 

\begin{deluxetable}{ccccrrr}
\tabletypesize{\scriptsize}
%\rotate
\tablecaption{Observed and de-reddened fluxes.}
\tablewidth{0pt}
\tablehead{
\colhead{Id} & \colhead{c({H$\beta$})} & \colhead{Ion} & \colhead{$\lambda$ (\AA)} 
& \colhead{F$_{\lambda}$} & \colhead{$\Delta$(F$_{\lambda}$)} & \colhead{I$_{\lambda}$} \\
(1) & (2) & (3) & (4) & (5) & (6) & (7) 
}
\startdata
PN001 &0.323$\pm$0.026 & HI    &     4340 &  47.6 &     3.7 & 52.4\\
      && HeII  &     4686 					       &   	58.0 &     4.4  &  60.1 \\      
      && HI    &     4861                                             &      100.0&     4.8  & 100.0 \\  
      &&[OIII] &     4959 					      &    	304.2&     7.4  &  298.9\\      
      &&[OIII] &     5007 					      &    	917.8&     12.2 &  894.0\\      
      &&HI     &     6563 					      &      362.2&     8.0  &  289.0\\  
\hline
PN002 & 0.343$\pm$0.030& HI    &     4340 &      42.3 &     3.9 & 46.9\\ 
      && HI    &     4861 				&     100.0 &     4.8        &  100.0\\  
      &&[OIII] &     4959 				&     304.1 &     7.5        &  298.5\\      
      &&[OIII] &     5007 				&     953.0 &     12.        &  927.04\\      
      && HI    &     6563 				&     368.3 &      8.        &  289.70  \\
\hline
PN003 & 0.460$\pm$0.014 &[OII]  &    3727  &     101.8  &     1.6  &  137.2\\
       &&HI     &    3835  					&       7.7  &     0.8          &   10.1\\     
       &&HeI    &    3889  					&     	 9.1  &     0.8          &   11.7\\      
       &&[NeIII]/HI &3968  					&       7.6  &     0.6          &    9.6\\     
       && HI    &    4100  					&       17.3 &     0.8          &   21.1\\ 
       && HI    &    4340  					&       37.5 &     1.2          &   43.1\\ 
       &&[OIII] &    4363  					&        7.4 &     0.7         &     8.5\\      
       && HeI   &    4471  					&        4.4 &     0.7         &     4.9\\     
       && HI    &    4861  					&       100.0&     1.8         &   100.0\\ 
       &&[OIII] &    4959  					&       178.5&     2.0         &   174.2\\     
       &&[OIII] &    5007  					&       540.1&     3.0         &   520.2\\     
       &&[NII]  &    5755  					&    	   3.7&     0.7         &     3.0\\     
       && HeI   &    5876  					&        22.4&     1.0         &    18.0\\      
       &&[NII]  &    6548  					&        43.5&     1.0         &    31.6\\     
       && HI    &    6563  					&       384.8&     2.4         &   280.0\\ 
       &&[NII]  &    6584  					&    	 128.2&     1.6         &    92.7\\     
       && HeI   &    6678  					&         5.2&     0.7         &     3.7\\     
       &&[SII]  &    6717  					&         7.2&     0.6         &     5.1\\     
       &&[SII]  &    6731  					&        10.3&     0.8         &     7.3\\     
       &&HeI    &    7065  					&         9.7&     0.8         &     6.5\\     
       &&[ArIII]&    7135  					&        14.8&     0.8         &     9.9\\     
\enddata
\label{tab_flux}
\tablecomments{(1) PN name; (2) nebular
extinction coefficient
\cbeta\ with its error;  (3) emitting ion; (4) rest-frame wavelength 
in \AA;  (5) measured line fluxes; (6) absolute errors on the measured line fluxes; 
(7) extinction corrected line fluxes. Both F$_{\lambda}$ (5)
and I$_{\lambda}$ (7) are expressed on a scale where  \hb=100. Table
\ref{tab_flux} is published in its entirety in the  
electronic edition of the {\it Astrophysical Journal}. A portion is 
shown here for guidance regarding its form and content.}
\end{deluxetable}

In order to confirm the goodness of our spectroscopic calibration we
compare our emission-line flux measurements of both \hii\ regions and
PNe with previously published fluxes of the same objects.  In Figure 1
we plot our measured fluxes against the ones from the literature,
where the two dashed lines mark differences of $\pm$0.15 dex between
the two sets.  We found a good agreement between the
sets, especially for bright emission lines, giving us confidence of a
sound spectral calibration.

In Figure~\ref{Fig_spectra} we show, as an example of the data quality, the
spectra of three PNe with different excitation.    

\begin{figure}
\resizebox{\hsize}{!}{\includegraphics[angle=0]{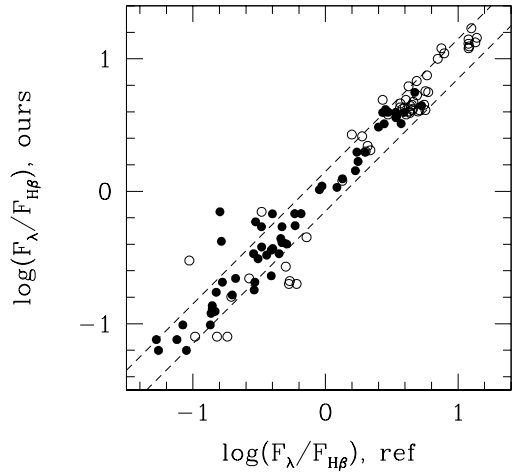}}
\caption{
Emission-line flux measurements of both \hii\ regions (filled
circles) and PNe (empty circles) are compared to previously published
fluxes. The fluxes of the \hii\ regions have been taken from  Kwitter \&
Aller~\cite{kwitter81};  
Vilchez et al.~\cite{vilchez88}; Crockett et al.~\cite{crockett06}, M07a; RS08, 
and the PN ones are from  Magrini et al.~\cite{magrini03}.}
\label{Fig_comp}
\end{figure}

\begin{figure*}
\resizebox{\hsize}{!}{\includegraphics[scale=2.3,angle=0]{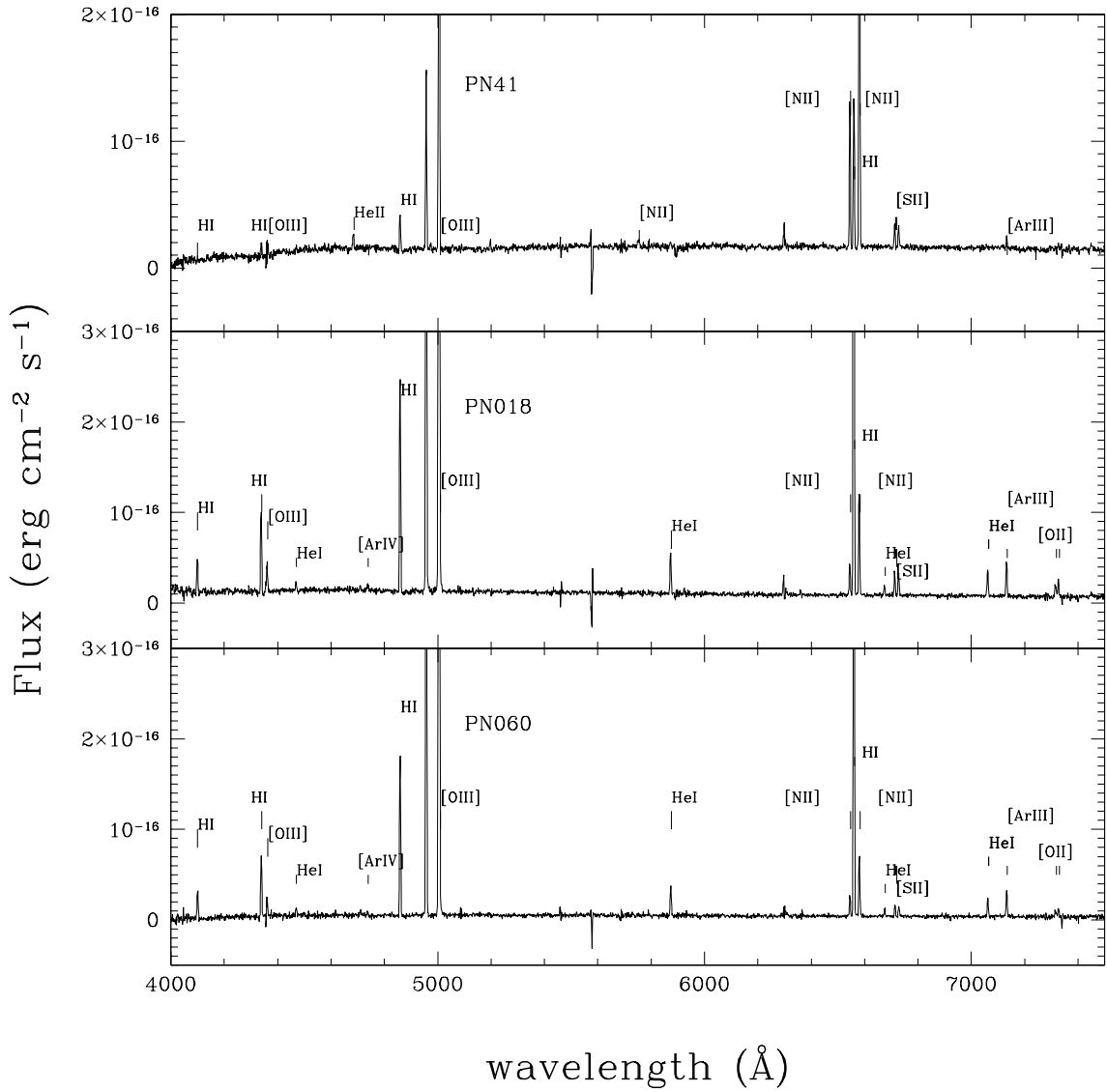}}
\caption{Spectra of three PNe with different excitation selected from our
  sample in order to show the data quality.}
\label{Fig_spectra}
\end{figure*}

\section{Plasma Diagnostics and Abundances}
\label{sect_abu}

\subsection{Electron Densities and Temperatures}
\label{sec_temp}
We have used the extinction-corrected intensities to obtain the PN electron
densities and temperatures. In order to calculate the electronic densities we
used the doublet of the sulfur lines
\sii$\lambda\lambda$6716,6731, while for the electron temperatures we 
used the ratios \oiii $\lambda$4363/($\lambda$5007 + $\lambda$4959) and \nii
$\lambda$5755/($\lambda$6548 + 
$\lambda$6584).  We performed 
plasma diagnostics by using the 5-level atom model included in the {\it nebular}
analysis package in IRAF/STSDAS (Shaw \& Dufour~1994). 

We have also used the
standard forbidden line diagnostics IRAF routines in {\it
nebular} to determine the electron temperatures. We have determined the low- and
medium-excitation temperatures   
from the \oiii\ and the \nii\ line ratios, respectively (see also
Osterbrock \& Ferland 2006, $\S$5.2). The temperature uncertainties
have been estimated by formal error propagation of the absolute
errors on the line fluxes (see also Table 4). The average
relative uncertainties in the determination of \teoiii\ and
\tenii\ are of the order of 5$\%$ and 15$\%$, respectively.

The \oiii\ and \nii\ temperature diagnostics are available, respectively,
for 32 and 8 PNe of our sample. Only for six PNe we can use both temperature
diagnostics. In order to improve the size of the PN sample with temperature
determinations we looked for correlations
between the electron temperature and various diagnostics of the nebular
excitation, similarly to the approach followed by Kaler~\cite{kaler86} for a
Galactic PN sample.

\begin{figure}
\resizebox{\hsize}{!}{\includegraphics[angle=0]{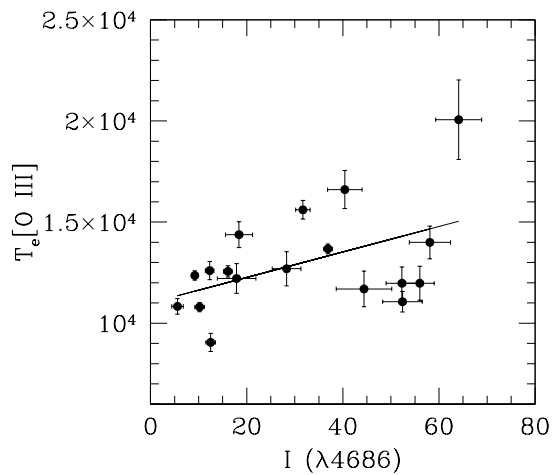}}
\caption{\oiii\ electron temperature vs. I$_{\lambda4686}$ line intensity,
  scaled for I$_{\rm H\beta}$=100 and 
corrected for extinction. The solid line is the fit done with the {\it fitexy} routine, shown in Eq.~(1). }
\label{Fig_3}
\end{figure}

In Figure~\ref{Fig_3} we plot the \oiii\ electron temperature against
the intensity of the I$_{\lambda4686}$ line, scaled for I$_{\rm H\beta}$=100
and corrected for extinction. Both quantities are plotted with
their formal error bars. Planetary nebulae in this plot are those hot
enough for helium to be doubly ionized. We find a clear correlation
between the  \oiii\ electron temperature and
the intensity of the I$_{\lambda4686}$ line, an effect that is expected as the 
central stars heat the medium-high excitation nebulae. In order to
use this correlation in our temperature calculation we have applied the routine
{\it fitexy} in Numerical Recipes (Press et 
al. 1992) to fit the relation between the two variables, taking into account
their errors, and minimizing $\chi^2$.

For I$_{\lambda4686}>$0 we find:
$${\rm T_e~[O~III]}=(63\pm8)~{\rm I_{\lambda4686}}+(11,000\pm200),\eqno(1)$$
where \teoiii\ is given in K and I$_{\lambda4686}$ is scaled to 
\hb=100. Note that using a different fitting routine, such as {\em sixlin}
(Isobe 
et al. 1990), we determine temperatures 
that agree with the ones of Eq. (1) at the 1\% level. We infer that we are
able to estimate the 
temperatures to better than 10\% with the above formulation.
 
If I$_{\lambda4686}$=0 we assume a constant value of T$_{\rm e}$[O
III], as in Kaler~\cite{kaler86}. The average T$_{\rm e}$[O III] for
I$_{\lambda4686}$=0 is 12100 $\pm$1500 K, which is compatible with the
continuity of Eq. (1). Note that both this value and Eq. (1) give
temperatures that are consistently higher than those found by
Kaler~\cite{kaler86}  
in Galactic PNe, an effect that is probably due to the lower
metallicity that characterizes the M33 stellar populations (see Stanghellini
et al.~\cite{stanghellini03}  
for a discussion on how metallicity affects the nebular temperatures).

\begin{figure}
\resizebox{\hsize}{!}{\includegraphics[angle=0]{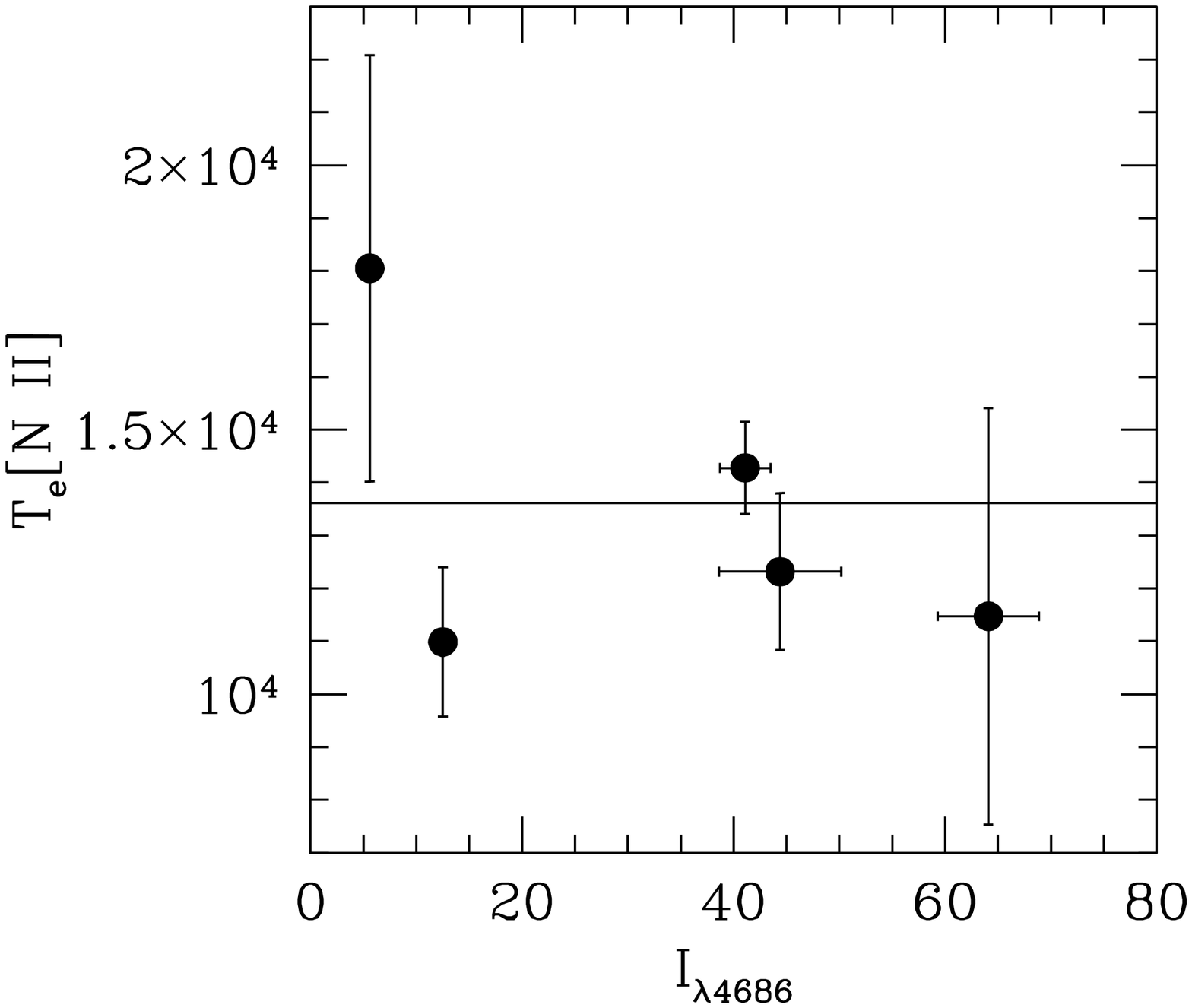}}
\caption{Same as Figure~3 but for the \tenii\ electron temperature.  The solid line
represents the mean value of \tenii= 13600 $\pm$2100 K, which has been
derived using the relative temperature errors as
weights. }
\label{Fig_2}
\end{figure}

In Figure~\ref{Fig_2} we show \tenii\ against I$_{\lambda4686}$, where
both quantities are plotted with their formal errors. The line
represents the mean value of \tenii= 13600 $\pm$2100 K, which has been
derived using the (inverse squared) relative temperature errors as
weights. Once again, while the physical behavior of the temperatures
is similar to that found in Kaler~\cite{kaler86}, the actual values are higher,
possibly reflecting the lower efficiency of the cooling process at low
metallicity. For I$_{\lambda4686} >$0 we adopt
\tenii=13,600 K.

If I$_{\lambda4686}$=0, we expect \tenii~to decrease with the stellar
temperature. The only way to frame this decline is to use the ratio
between the fluxes of the singly-ionized to the doubly-ionized oxygen. By
correlating the oxygen line strengths, corrected for extinction, to
the electron temperatures for the PNe with measured \nii ~lines 
we find:

$$  {\rm T_e~[N~II]}=(11,700\pm2800)-(2800\pm4000)~{\rm log}{{{\rm
      I_{\lambda\lambda3727-29}}}\over{{\rm I_{\lambda 4959}}+{\rm
      I_{\lambda5007}}}}. \eqno(2)$$ 

Note that in  our spectra the $\lambda\lambda3727-29$ lines
are always blended.  If the oxygen lines are not available, and
I$_{\lambda4686}$=0 we assume
\tenii=14,100 $\pm$2800 K, which, in this case, is the weighted mean of the
temperatures.  

\subsection{Chemical abundances}

\begin{deluxetable}{cll}
\tabletypesize{\scriptsize}
%\rotate
\label{tab_allabu}
\tablecaption{Plasma Diagnostics and Abundances
}
\tablewidth{0pt}
\tablehead{
\colhead{Id} & \colhead{} & \colhead{}\\
(1) & (2) & (3)}
\startdata
PN001          &         &  \\     
               &Te(OIII)  &  14650\tablenotemark{b} \\  
               &Te(NII)   &  13600\tablenotemark{b} \\  
               &HeII/H         &  0.063 \\ 
               &He/H           &  0.063 \\
               &OIII/H         &  7.586 10$^{-05}$\\ 
               &ICF(O)         &  1.000 \\
               &O/H            &  7.586 10$^{-05}$\\ 
\hline
PN002         &         &  \\    
               &Te(OIII)  &  14650\tablenotemark{b} \\  
               &Te(NII)   &  13600\tablenotemark{b} \\  
               &OIII/H         &  7.884 10$^{-05}$\\ 
               &ICF(O)         &  1.000 \\
               &O/H            &  7.884 10$^{-05}$\\ 
\hline
PN003          &         &  \\                  
               &Te(OIII)       &  14000\tablenotemark{a}    \\   
               &Te(NII)        &  15200\tablenotemark{a}    \\
               &Te(OIII)  &  12060\tablenotemark{b}   \\  
               &Te(NII)   &  14120\tablenotemark{b}   \\  
               &Ne(SII)        &  2540    \\   
               &HeI/H          &  0.125   \\
               &He/H           &  0.125   \\ 
               &OII/H          &  1.590 10$^{-05}$\\ 
               &OIII/H         &  6.500 10$^{-05}$\\ 
               &ICF(O)         &  1.000 \\ 
               &O/H            &  8.090 10$^{-05}$\\ 
               &NII/H          &  7.168 10$^{-06}$\\  
               &ICF(N)         &  5.087\\ 
               &N/H            &  3.646 10$^{-05}$\\ 
               &ArIII/H        &  4.47 10$^{-07}$\\  
               &ICF(Ar)        &  1.245 \\ 
               &Ar/H           &  8.359 10$^{-07}$\\  
               &SII/H          &  1.736 10$^{-07}$\\  
               &ICF(S)         &  1.276  \\ 
               &S/H            &  1.665 10$^{-06}$\\  
\enddata
\tablenotetext{a}{computed from electron temperature diagnostic lines. }
\tablenotetext{b}{derived from the relations in $\S$3.}
\tablecomments{(1) identification name; (2) label of each plasma diagnostic and abundances available;  (3) 
relative value obtained from our analysis.
Table~3 is published in its entirety in the 
electronic edition of the {\it Astrophysical Journal}. A portion is 
shown here for guidance regarding its form and content.}
\end{deluxetable}

We computed the ionic abundances using the {\it nebular} analysis
package in IRAF/STSDAS (Shaw \& Dufour 1994).  The elemental
abundances were then determined by applying the ionization correction
factors (ICFs) following the prescriptions by Kingsburgh \&
Barlow~\cite{kb94} for the case where only optical lines are
available. As discussed in the previous $\S$, the \oiii\ ~$\lambda$4363 emission
line was measured with a sufficiently high signal to noise ratio in
32 PNe, affording the direct determination of the \teoiii. We could also
directly measure \tenii~ in 8 PNe.  For the other targets we adopted
the temperatures derived from the relations we have described in the
previous Section.  In the abundance analysis we have used
\tenii\ for the calculation of the N$^+$, O$^+$, S$^+$ abundances, while
\teoiii\ was used for the abundances of O$^{2+}$, S$^{2+}$, Ar$^{2+}$,
He${^+}$, and He$^{2+}$. 

We calculated the abundances of \hei\ and \heii\ using the
equations 
of Benjamin et al.~\cite{benjamin99} in two density regimes, i.e. \ne\
$>$1000 cm$^{-3}$ and $\leq$1000 cm$^{-3}$.  The Clegg's collisional
populations were taken into account (Clegg 1987).  Due to the effect
of double collisions not being properly corrected, the He$^+$/H ionic
abundance from the $\lambda$7065 emission-line is quite different
from the ionic abundance computed from $\lambda\lambda$4471, 5876, and
6678. We thus omit the abundance derived from the former line from the
average.
The computed individual ionic and elemental abundances of each PN are shown
in Table 3, where column (1) gives the PN identification name; columns (2) and
(3) show the plasma diagnostic and abundances available for each PN, where in
column (2) we label each diagnostic, and in column (3) we give the
relative value obtained from our analysis.

\begin{deluxetable}{lllllllll}
\tabletypesize{\scriptsize}
%\rotate
\tablecaption{Typical Errors in dex}
\tablewidth{0pt}
\tablehead{
\colhead{Magnitude} & \colhead{T$_e$} & \colhead{$\Delta$(He/H)} & \colhead{$\Delta$(O/H)}& \colhead{$\Delta$(N/H)}& \colhead{$\Delta$(Ne/H)}& \colhead{$\Delta$(Ar/H)}& \colhead{$\Delta$(S/H)} & \colhead{N. PNe}  \\
(1) & (2) & (3) & (4) & (5) & (6) & (7) & (8) & (9)
}
\startdata
20.60-21.00 &   l   &  0.03  & 0.025     & 0.06   & 0.05   & 0.06   & 0.09   & 8\\
            &   c   &  -     & 0.17      & 0.40   & 0.43   & 0.34   & 0.43   & 1\\
21.01-21.50 &   l   &  0.04  & 0.03     & 0.06   & 0.06   & 0.11   & 0.11   & 9\\
            &    c  &  0.09  & 0.20     & 0.34   & 0.43   & 0.30   & 0.30   & 8 \\
21.51-22.00 &   l  &   0.04  & 0.03     & 0.04   & 0.08   & 0.09   & 0.09   & 5 \\
            &   c   &  0.08  & 0.21     & 0.34   & 0.43   & 0.34   & 0.30   & 8 \\
22.01-22.50 &   l   &  0.08  & 0.04    & 0.07   & 0.15   & 0.17   & 0.11   & 6\\
            &   c   &  0.11  & 0.21    & 0.40   & 0.43   & 0.40   & 0.30   & 16 \\
22.51-23.00 &   l   &  0.13  & 0.09    & 0.16   & 0.16   & 0.17   & 0.11   & 4\\
            &   c   &  0.17  & 0.20    & 0.40   & 0.43   & 0.43   & 0.30   & 9 \\
23.51-24.00 &   l   &  0.09  & 0.07    & 0.09  & 0.13  & 0.18   & 0.16   & 1\\
            &   c   &  0.21  & 0.23    & 0.40   & 0.43  & 0.43   & 0.30   & 12 \\
24.01-24.50 &   c   &  0.11  & 0.21    & 0.34  & 0.43  & 0.43   & 0.43   & 5 \\
24.51-25.00 &   c   &  -  & 0.45    & -  & -  & -  & -   & 2 \\
25.01-25.50 &   c   &  -  & 0.5    & -  & -  & -  & -   & 1 \\                           
\enddata
\label{Tab_error}
\tablecomments{(1) Magnitude
  range; (2) method of the electron temperature determination,
where {\it l} means that the temperature has been measured based on
an emission line and {\it c} means that we used the correlations
described in $\S$3; (3-8) typical errors in dex of the total abundances; (9) number of PNe available in each bin.}
\end{deluxetable}

The formal errors on the ionic and total abundances were computed taking
into account the uncertainties in the observed fluxes, in the electron
temperatures and densities, and in \cbeta.  The errors (in dex) of the final
abundances are given in Table~\ref{Tab_error}.  Typical errors are
given for PNe in different \oiii\ magnitude ranges, taking also into
account whether the electron temperatures were derived from
emission-line diagnostic or from the relations
described in $\S$~\ref{sec_temp}. In the latter case, we assumed a
percentage error on the electron temperature of $\sim$10\%, which is
the dispersion of the relation used to derive the temperature.  The
first column of Table~\ref{Tab_error} gives the magnitude range;
column (2) gives the method used for the electron temperature determination,
where {\it l} means that the temperature has been measured based on
an emission line and {\it c} means that we used the correlations
described in $\S$3; columns (3)
to (8) give the typical errors in dex of the total abundances. The
last column gives the number of PNe on each bin.

\begin{deluxetable}{ccccccccc}
\tabletypesize{\scriptsize}
\rotate
\tablecaption{Average chemical abundances and abundance ratios}
\tablewidth{0pt}
\tablehead{
\colhead{Sample} & \colhead{He/H} & \colhead{O/H (10$^{-4}$)} & \colhead{N/H (10$^{-4}$)} & \colhead{Ne/H (10$^{-5}$)} & \colhead{Ar/H (10$^{-6}$) } & \colhead{S/H (10$^{-6}$)} & \colhead{N/O} & \colhead{Ne/O} \\
(1) & (2) & (3) & (4) & (5) & (6) & (7) & (8) & (9)
}
\startdata
M33 PNe, Disk (91)            & 0.118$\pm$0.074 & 1.96$\pm$1.27 & 1.09$\pm$1.54 & 4.09$\pm$3.08 & 1.19$\pm$0.57 & 3.34$\pm$1.71 & 0.40$\pm$0.32 & 0.17$\pm$0.06 \\ 
M33 PNe, non-Type~I (72) & 0.114$\pm$0.083 & 1.73$\pm$0.93 & 0.39$\pm$0.19 & 3.65$\pm$2.11 & 1.12$\pm$0.50 & 4.46$\pm$2.30 & 0.18$\pm$0.06 & 0.17$\pm$0.06 \\ 
M33 PNe, Type~I (19)        & 0.135$\pm$0.027 & 2.70$\pm$1.88 & 1.66$\pm$1.88 & 4.83$\pm$4.25 & 1.38$\pm$0.68 & 6.64$\pm$3.80 & 0.58$\pm$0.33 & 0.16$\pm$0.04 \\ 
M33 PNe, Halo (2)              & 0.120:                   & 2.62$\pm$0.31 & 0.84:                 & 3.47$\pm$0.82 & 1.24$\pm$0.28 & 6.45:                   & 0.31:                & 0.13$\pm$0.03 \\
\hline
M33 HII regions   & 0.101$\pm$0.015(a) & 2.04$\pm$0.75(b) & 1.28$\pm$0.57(a) & 4.24$\pm$2.61(c) & 1.31 $\pm$0.45(a) & 5.85$\pm$2.28(a) & 0.06$\pm$0.02(a)&0.20$\pm$0.06(c)            \\

\hline 
Solar value(d)      & 0.085$\pm$0.02  & 4.57$\pm$0.04 & 0.60$\pm$0.09 & 6.9$\pm$1.0   & 1.51$\pm$0.30 & 13.8$\pm$2.0  & 0.13$\pm$0.10 & 0.15$\pm$0.05\\
\hline
LMC PNe         &	0.103$\pm$0.026 &2.32$\pm$1.65  & 1.48$\pm$1.75 &  4.04$\pm$3.60& 1.14$\pm$0.72 & 3.46$\pm$8.88 & 0.87$\pm$1.15 & 0.17 $\pm$0.09\\
SMC PNe       &	0.113$\pm$0.022 &1.05$\pm$0.46  & 0.28$\pm$0.33 &  1.77$\pm$1.32& 0.59$\pm$0.59 & 4.80$\pm$6.57 & 0.28$\pm$0.50 & 0.17 $\pm$0.08\\
Galactic PNe          &	         0.123$\pm$0.042       &  3.53$\pm$1.95             &  2.44$\pm$3.46             & 9.68$\pm$7.98              &        1.26$\pm$1.24       &  \dots             &
     0.67$\pm$0.82         &      0.25$\pm$0.10        \\
\hline 
\enddata
\tablecomments{(1) the PN sample; (2-7) average chemical abundances by number with their rms uncertainties; (8-9) $<$N/O$>$ and $<$Ne/O$>$, the average of N/O and Ne/O values for each object. 
 }
\tablenotetext{a}{Chemical abundances computed using the sample of M07a, extended to all literature data with direct electron temperature measurement. }
\tablenotetext{b}{O/H is calculated with the cumulative sample of M07a, which includes also all previous O/H with direct electron temperature measurement, and RS08.}
\tablenotetext{c}{Ne/H and Ne/O are from Crockett et al.~\cite{crockett06}.}
\tablenotetext{d}{Solar value from Asplund et al.~\cite{asplund05}.}
\tablenotetext{:}{Value from a single object.}
\label{tab_avabu}
\end{deluxetable}

\section{The PN population}
\label{sect_pne}

Most of the PNe discovered in M33 belong to its disk, with only two PNe
having radial velocities compatible with the halo population (Ciardullo et
al.~2004). Therefore, we have the opportunity to analyze in detail a pure
disk PN population by excluding from our analysis the two suspected halo
PNe (namely PN067 and PN024).  We have also excluded from our analysis
the PN039 since its physical conditions do not allow us to obtain
abundance diagnostics. PN039 has a WR-nucleus and will
be discussed in detail in the Appendix. 

\subsection{The He/H vs N/O diagram}
The plot of the N/O ratio vs the He/H one is a classical diagnostic diagram
used to discriminate PNe of different 
types, i.e. with different mass progenitors. The N/O ratio provides
information about the stellar 
nucleosynthesis during the AGB phase of LIMS. On the one hand, nitrogen is
produced in AGB stars in two ways: by neutron capture, during the CNO
cycle, and by hot-bottom burning. Hot-bottom burning
produces primarily nitrogen but occurs only if the base of the convective
envelope of the AGB stars is hot enough to favor the conversion of
$^{12}$C into $^{14}$N.  Thus nitrogen is expected to be mostly enriched
in those PNe with the most massive progenitors, i.e. with turnoff mass
larger than $\sim$3 M$_{\odot}$  (van der Hoek \& Groenewegen 1997;
Marigo 2001). The He/H ratio gives also an indication of the initial mass of the
progenitor, the nebula is enriched progressively for more massive stars,  it
reaches a plateau between 3 and 4 M${_\odot}$, and then 
it increases again toward the higher masses (Marigo 2001).

\begin{figure}
\resizebox{\hsize}{!}{\includegraphics[angle=0,scale=1.00]{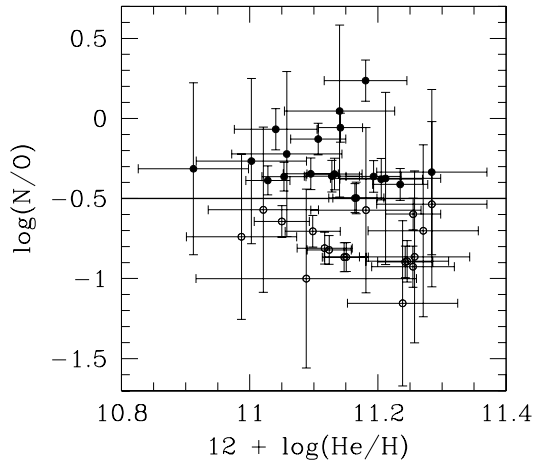}}
\caption{
$\log$(N/O) versus 12+$\log$(He/H). 
Type I PNe and non-type I PNe are plotted as filled circles and 
empty circles respectively, according to the definition of
Dopita~\cite{dopita91}. 
The continuous line mark log(N/O)$>$-0.5 that, independently of helium
abundance, defines  
the area of Type I PNe.}
\label{Fig_heno}
\end{figure}

In order to make a selection of the PNe with high-mass progenitors, in
Figure~\ref{Fig_heno}, we plot the M33 disk PNe on the He-N/O plane.
Peimbert \& Torres-Peimbert \cite{peimbert83} found that Galactic PNe
that were nitrogen- and helium-enriched also had bipolar shape, and
were located closer to the Galactic plane. It looked like these PNe
formed a younger population, which they defined as those with
$\log~$(N/O$)\ge -0.3$ and He/H$\ge\sim$0.125, and called them Type I
PNe.  Kingsburgh \& Barlow~\cite{kb94} analyzed a much larger sample and
redefine the Galactic Type I PNe as those having log(N/O)$\ge$-0.1,
with 18$\%$ of the PNe in their sample being of Type I. Dopita~\cite{dopita91}
analyzed LMC PNe and noted that the Type I definition needed a
revision to allow for the lower metallicity of the LMC with respect to
the Galaxy, setting the limit of Type I PNe to log(N/O)$\ge$-0.5 for
the LMC. Leisy \& Dennefeld~\cite{leisy06} basically confirmed this
limit. Magrini et al.~\cite{magrini04} also noted that the definition
of Type I PNe depends on the metallicity, because the amount of
nitrogen that can be produced by hot bottom burning is dependent on
the amount of carbon present, and also because the oxygen abundance
depends on the metallicity of the galaxy.  Since the oxygen
metallicity of M33 is very similar to that of the LMC, we use the Type
I limits as in Dopita~\cite{dopita91}, namely, Type I PNe are those with
log(N/O)$>$-0.5, independent of helium abundance.
With this definition we found 19 PNe being Type I in the disk of M33 ($\sim$20\% of the whole disk population) 
and one in the halo.

\subsection{The Ne/H and S/H vs O/H diagrams}

The study of the chemical evolution of galaxies needs strong
constraints regarding the past composition of the ISM, and PNe can supply
them, in particular with their soundly-determined oxygen abundances.  
The questions is whether oxygen (and neon) is modified
in LIMS while in their AGB phase.

A possible way to verify this is to study the relationship between neon
and oxygen.  These elements derive both from primary
nucleosynthesis, mostly in stars with M$>$10 M$_{\odot}$.  If the O/H
and Ne/H abundances are
really independent of the evolution of the PN progenitors through the
AGB phase, a tight correlation between their abundance should be observed.

\begin{figure}
\resizebox{\hsize}{!}{\includegraphics[angle=0,scale=1.00]{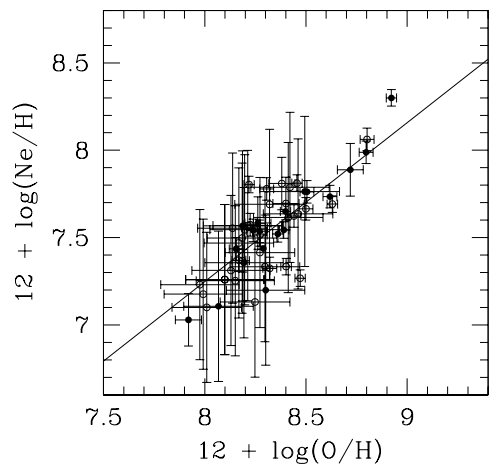}}
\caption{The relationship between oxygen and neon abundances. Symbols are as in Figure~\ref{Fig_heno}.
The continuous line is the weighted least square fit to the complete sample of PNe.}
\label{Fig_oxyne}
\end{figure}

In Figure~\ref{Fig_oxyne} we show O/H against Ne/H for the 55 M33 PNe where both
abundances
were available.  The
slope of the correlation is close to unity, 0.90$\pm$0.11, with a
correlation coefficient R$_{\rm xy}$=0.81, pointing at a locked variation of
these 
two elements.  Recent results from  Wang \&
Liu~\cite{wang08} indicate that oxygen and neon could be manufactured
with similar yields also in LIMS, but only
at very low metallicities, 12 + $\log$(O/H) $<$ 8. These results are
however still not explained by current nucleosynthesis theories that
predict different channels, and thus different yields, for Ne and O
production in LIMS at low metallicity
(cf., e.g., Karakas \& Lattanzio 2003, Marigo et al. 2003). Our findings
confirm that in 
 M33, as  in LMC, and the Galaxy, there is no
evidence of enhancement of oxygen and neon in PNe.

\begin{figure}
\resizebox{\hsize}{!}{\includegraphics[angle=0,scale=1.0]{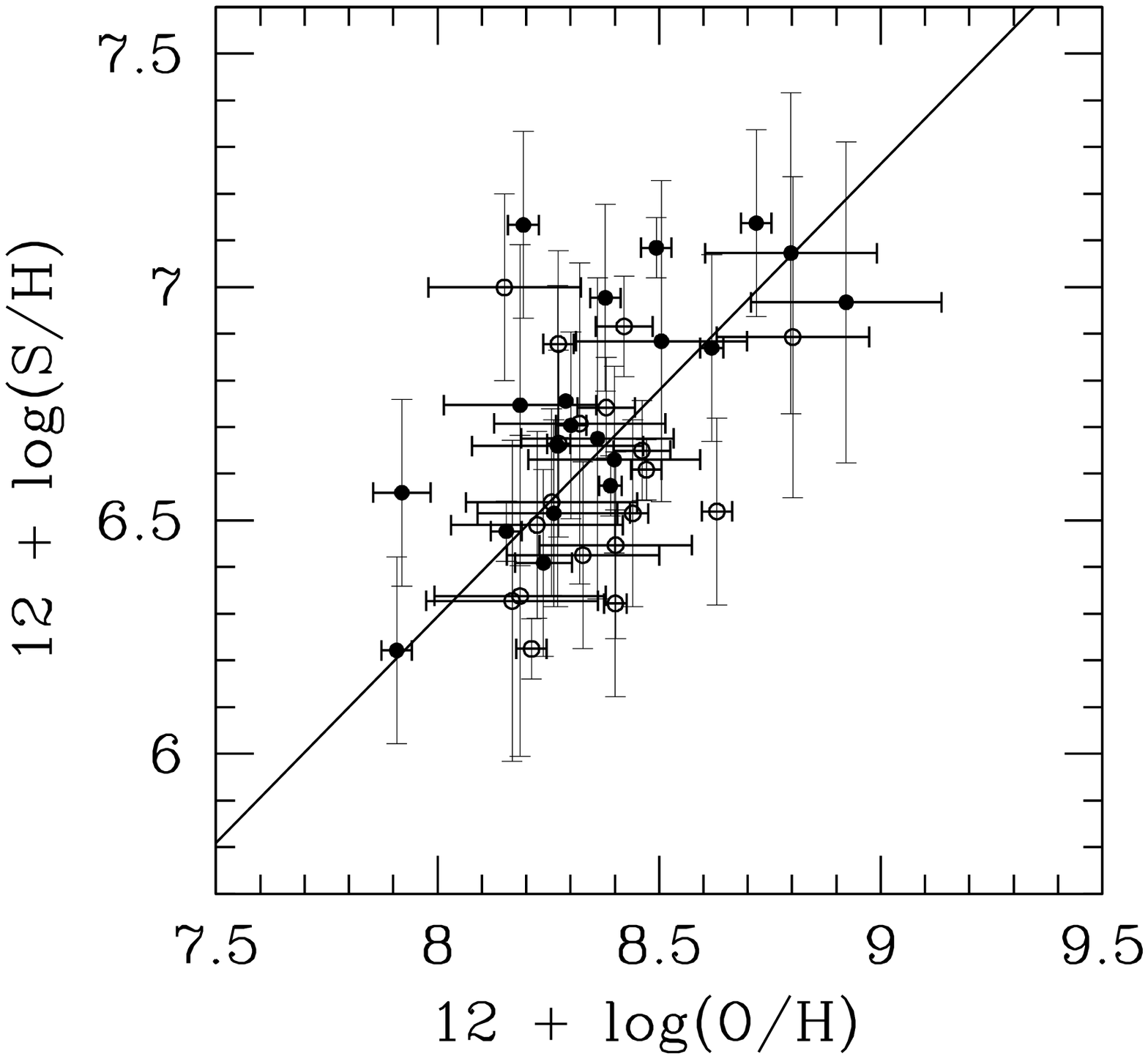}}
\caption{The relationship between oxygen and sulphur abundances. Symbols are as in Figure~\ref{Fig_heno}.
The continuous line is the weighted least square fit to the complete sample of PNe.}
\label{Fig_oxysu}
\end{figure}

Similar correlations are expected also for sulphur versus oxygen,
since sulphur is manufactured by massive stars.  In Figure
\ref{Fig_oxysu} we plot the sulfur versus oxygen abundances
of the 38 M33 PNe where these abundances are available.  Their relationship
has a slope of 0.97$\pm$0.22 and a 
correlation coefficient of 0.51, showing moderately good correlation.
We refrain from showing and analyzing the argon to oxygen relation. As seen
in Stanghellini et al.~\cite{stanghellini06},  
these quantities might not correlate due to their different origin. Even if
both elements 
are manufactured in massive stars, argon is synthesized in very different
$\alpha$ -element 
processes than  oxygen, thus a lockstep of these elements is not completely
expected.

\section{Comparison of abundances of different stellar populations, and of PN populations of other galaxies}
\label{sec_comp}

The average chemical abundances of our M33 PNe are reported in
Table~\ref{tab_avabu} 
where the first column describes the selected sample,
columns (2) to (7) give the average total abundance of He/H,
O/H, Ne/H, Ar/H, and S/H, whereas the last two columns, (8) and (9),
give the mean values of N/O and Ne/O.  
The average abundances were computed excluding
those abundances from upper limit detections. When the emission lines for
temperature diagnostics 
were not detected, we adopted the electron temperatures  discussed in
$\S$\ref{sec_temp}. Their larger uncertainties 
were taken into account by larger formal errors in the total chemical
abundances as described in Table \ref{Tab_error}. 

We group the PNe in M33 into four
populations:
{\em i}) the whole PN disk population; {\em ii})
the disk population excluding those PNe which are believed to be Type
I; {\em iii}) the Type I PNe, and {\em iv}) the halo PN
population as classified by Ciardullo et al.~\cite{ciardullo04}. For the sake of
comparison we add to the Table the average chemical abundances of the M33
\hii\ regions, the solar abundances, and the average abundances of the PNe in
LMC, SMC 
(Stanghellini 2008) and the Galaxy (Stanghellini et al. 2006).
Note, however, that the population of the Type I and non-Type I classes
remains somewhat uncertain, since about a dozen PNe 
could belong to either class based on their formal abundance errors.
However, the chemical abundances and distributions of Type I and non-Type I PNe in M33 are very similar, so that the displacement of a few
PNe from a group to the other  would not change our results significantly.
 
An inspection of Table 5 affords several interesting clues on the
chemical evolution of galaxies.  First, if we compare the average
abundances of the $\alpha$-elements of the M33 disk PNe with those of
the \hii\ regions, we do not note significant enrichment in the
\hii\ regions, leading to the conclusion that these elements are not
substantially modified in the LIMS evolution. We substantially found a
flat age-metallicity relationship in M33 all the way to the time of PN
formation.  
Second, we infer that the metallicity of the PNe in M33,
based on oxygen abundances, is sub-solar. If we consider the entire sample
of disk PNe in M33 we see that the averages of $\alpha$-elements are
virtually identical to those of PNe in the LMC, with PN metallicity
about 1/2 that of the Galaxy, and twice the SMC on average, both in
oxygen and neon.

From Table 5 we see that the average Ne/O ratio in the general
population of M33 PNe is identical to that of the LMC, but lower than
the mean Ne/O in Galactic PNe.  This finding is right on the mark with
the recent results of Wang \& Liu~\cite{wang08}. In fact Wang \&
Liu~\cite{wang08} found that the Ne/O ratio increases
with increasing oxygen metallicity both in PNe and in \hii\
regions. This suggests a different enrichment history of neon and
oxygen in the ISM and thus probably different production mechanisms of
these two $\alpha$-elements in massive stars, in agreement with
current theoretical calculations by Kobayashi et
al.~\cite{kobayashi06}.  While we should expect the solar Ne/O ratio
to be consistent with that of Galactic disk PNe and \hii\ regions, we
note that the Asplund et al.'s~\cite{asplund05} value, as well as
Grevesse \& Sauval's~\cite{grevesse98}, is slightly discrepant in this
scenario.  Both PNe and \hii\ regions in the Galactic disk give a
consistent Ne/O ratio of $\sim$0.25, higher than the solar value. Our
results on M33 PNe seem thus to be consistent with the recent
suggestions of a revision of the solar Ne/O ratio and the absolute
neon (e.g. Wang \& Liu~2008, Rubin et al.~2008).
 
The N/O average shown in Table~5 for M33 PNe is lower than the
Galactic and higher than the SMC values.  The comparison to the LMC
average has limited importance, giving the large range of LMC N/O
values.

\section{The abundance gradients}
\label{sect_grad}

\begin{figure*}
\resizebox{\hsize}{!}{\includegraphics[angle=0,scale=1.00]{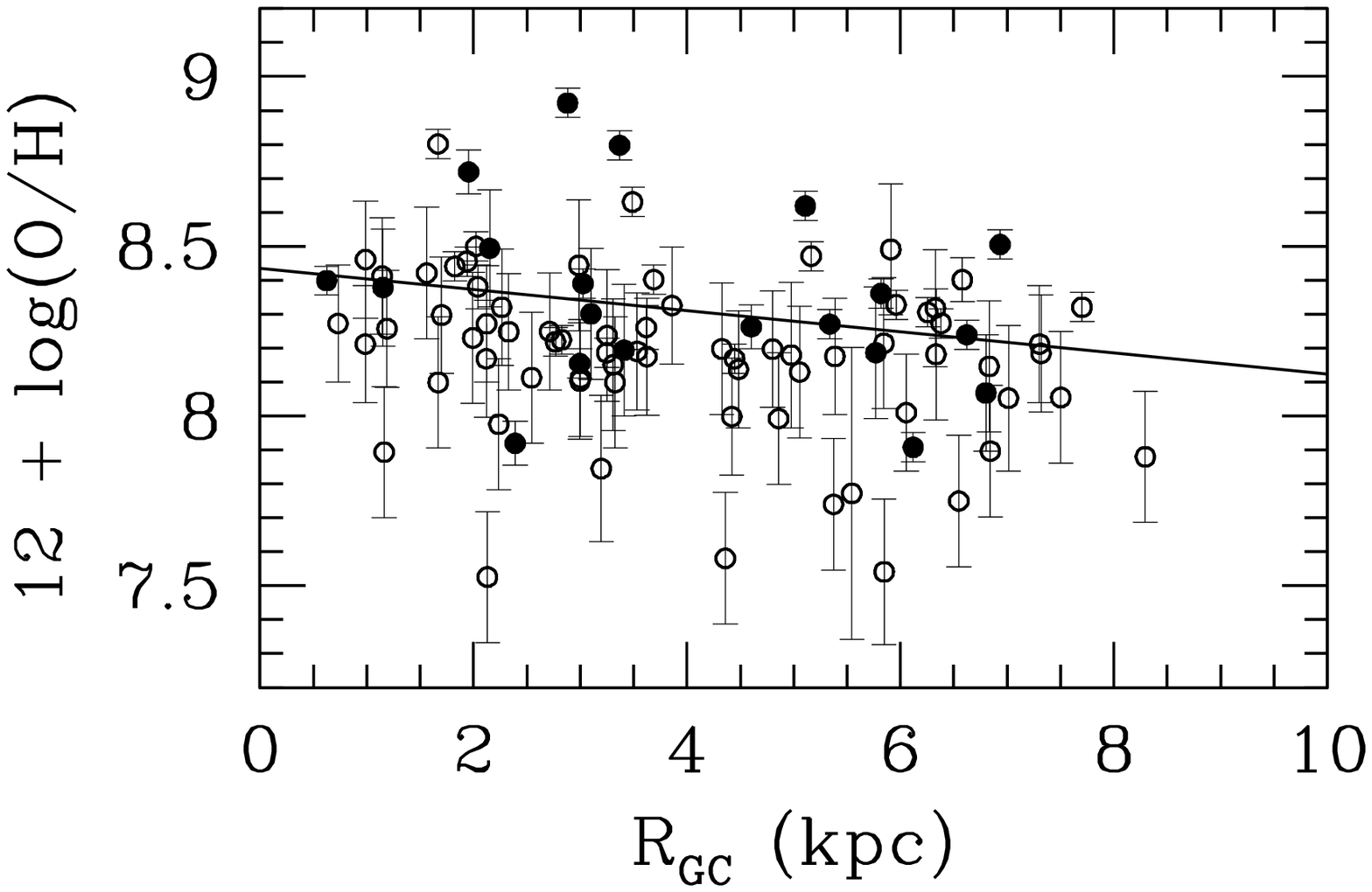}}
\caption{The radial gradient of oxygen abundance. Symbols are as in
  Fig.~\ref{Fig_heno}.  The continuous line is  
the weighted least square fit to the complete sample of disk PNe. Slopes and
zero-points are shown in Table~\ref{Tab_grad}.} 
\label{Fig_oxy}
\end{figure*}

\begin{figure}
\resizebox{\hsize}{!}{\includegraphics[angle=0,scale=1.0]{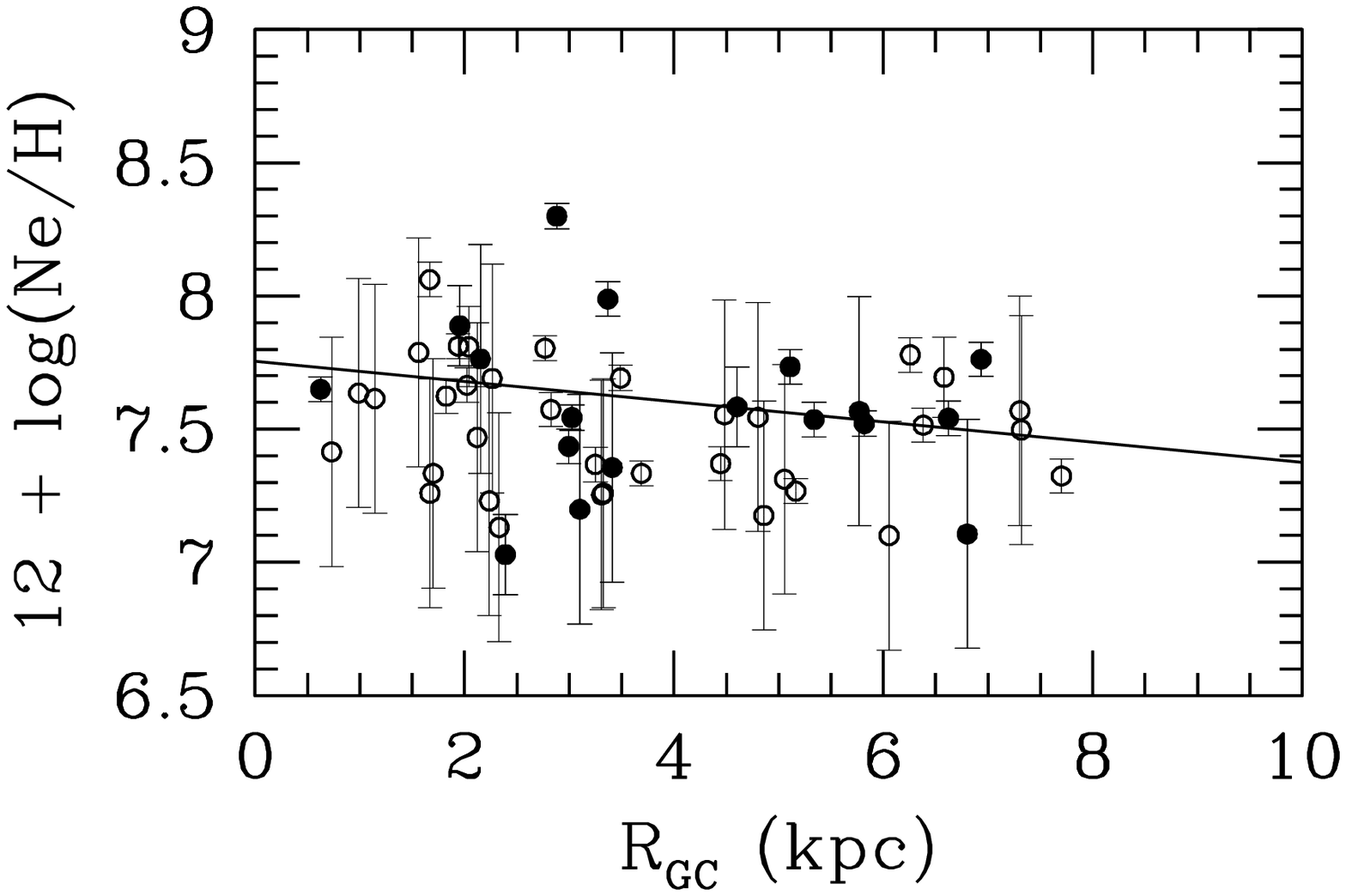}}
\caption{The radial gradient of neon abundance. Symbols are as in
  Fig.~\ref{Fig_heno}. The continuous line is  
the weighted least square fit to the complete sample of disk PNe. Slopes and zero-points are shown in Table~\ref{Tab_grad}.}
\label{Fig_ne}
\end{figure}

\begin{figure}
\resizebox{\hsize}{!}{\includegraphics[angle=0,scale=1.0]{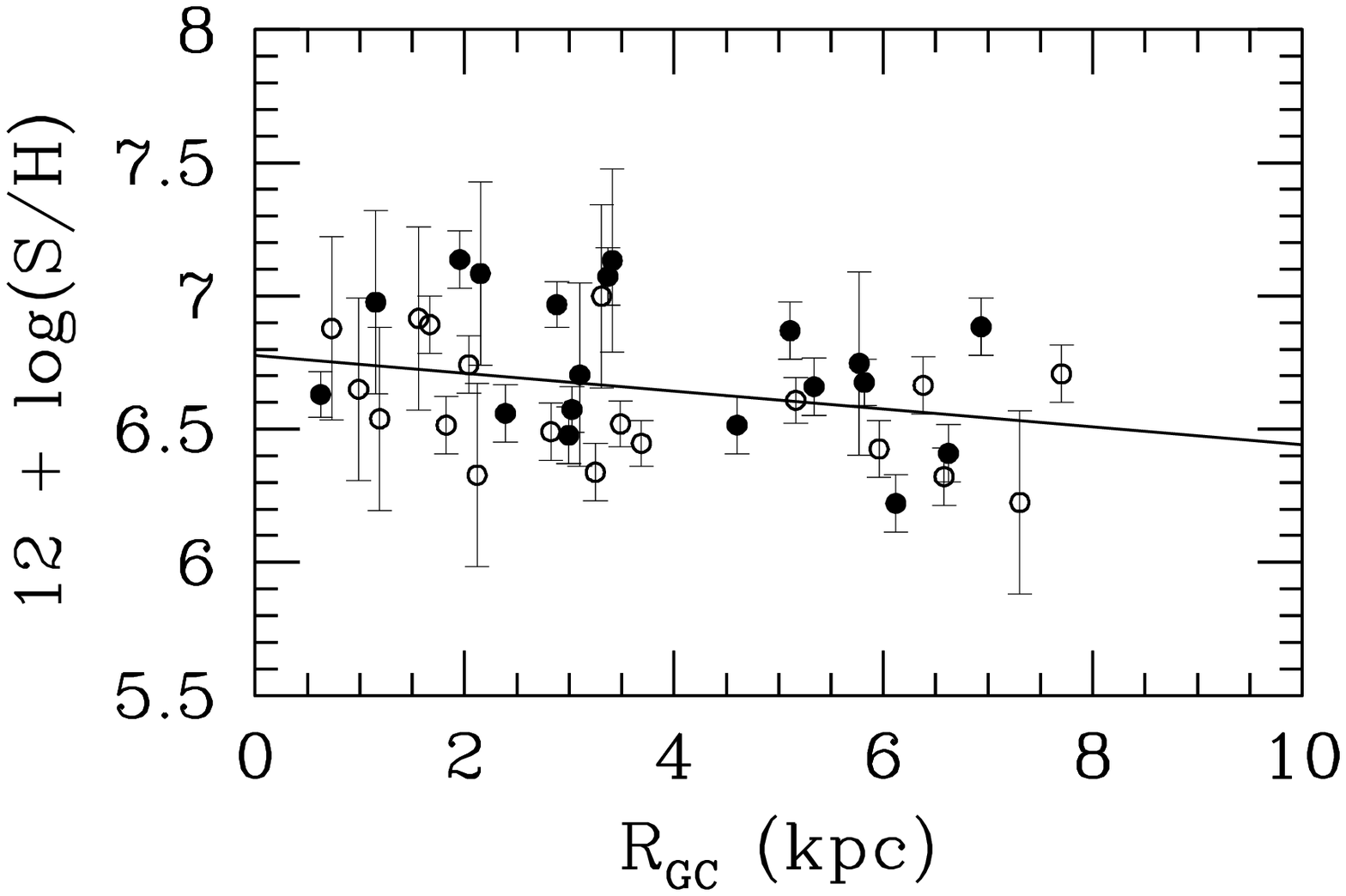}}
\caption{The radial gradient of sulphur abundance. Symbols are as in Fig.~\ref{Fig_heno}. The continuous line is 
the weighted least square fit to the complete sample of disk PNe. Slopes and zero-points are shown in Table~\ref{Tab_grad}.}
\label{Fig_su}
\end{figure}

In Figure~\ref{Fig_oxy}, we show the oxygen abundance as a function of
galactocentric distance for our sample of disk PNe. 
The galactocentric distances were computed by adopting a
distance to M33 840~kpc, as determined by Freedman et al.~\cite{freedman91}, 
an average and well-accepted distance estimate.
A weighted linear least-square fit to the
complete disk sample gives a gradient:

$$ 12 + {\rm log(O/H)} = -0.031 (\pm 0.013) ~  {\rm R_{GC}} + 8.44 (\pm
0.06),\eqno(3)$$ 

where R$_{\rm GC}$ is the de-projected galactocentric distance in kpc,
computed assuming an inclination of 53\arcdeg, and a position angle
of 22\arcdeg.  In Figures ~\ref{Fig_ne}, and
\ref{Fig_su} we plot the radial metallicity gradients of Ne/H and S/H.  
The solid lines are the weighted linear fits to the complete
sample of disk PNe (first row of Table~\ref{Tab_grad} for each element). In
all these 
figures the two symbols refer to Type I (filled circles) or non-Type I (empty
circles) PNe. 

In Table~\ref{Tab_grad} we show the slopes (column~2) and zero-points
(chemical abundances in the centre of the galaxy, column~3) of the
metallicity gradients considering different samples of PNe.  The sample
used and the number of PNe included are given in columns (4)
and (5) respectively. For each element, the first row report the gradient
obtained 
considering the whole sample of disk PNe (thus excluding the two
possible halo PNe), the second row gives the gradient computed using non-Type
I PNe,  
and the third row gives the gradient obtained for Type I PNe.

The gradients are computed with a weighted least mean square fit.  For
each element the slope of the three different samples are consistent
within the errors.  
%Only in the case of sulphur the gradient of
%non-Type I PNe is marginally consistent with the other gradients (all
%disk and Type I PNe), but the sample is very limited.  
Note that the
metallicity gradient of an old stellar population could be affected by
the radial migration of stars during cosmic times.  The present-time
location of a PN could not necessarily be the place where it was born.
Radial mixing of stars is believed to be due to several mechanisms,
among them the diffusion of stars on their orbits because of various
irregularities in the galactic potential (cf. Wielen et al. 1996), to
the passage of spiral patterns (cf. Minchev \& Quillen 2006), to
changes in the angular momentum due to non-asymmetric forces due to
molecular clouds (Spitzer \& Schwarzzchild 1953).  Upper limits to the
radial migration rate were estimated by De Simone et
al.~\cite{desimone04} and by Haywood~\cite{haywood08}, giving similar
values around 1.5-3.7 kpc Gyr$^{-1}$, limited to a radial region $\pm$2~kpc
from the birth place.  The model of Sellwood \&
Binney~\cite{sellwood02} show that spiral waves in galaxy disks churn
both the interstellar medium and the stars, affecting their metallicity
gradients.  The effect of stirring of the entire disk due to the
spiral waves is a flattening of the metallicity gradient.

However, the fact that metallicity gradients survive and are well
observed in disk galaxies seem to indicate that the effectiveness of
the migration processes is moderate.

\begin{deluxetable}{ccccc}
\tabletypesize{\scriptsize}
%\rotate
\tablecaption{Chemical abundance gradients}
\tablewidth{0pt}
\tablehead{
\colhead{Element} & \colhead{Slope} & \colhead{Zero} & \colhead{sample} & \colhead{\# PNe}\\
(1) & (2) & (3) & (4) & (5) 
}
\startdata
O/H               & -0.031$\pm$0.013 &  8.44$\pm$0.06    & All PNe                   & 91              \\
O/H               & -0.030$\pm$0.013 &  8.41$\pm$0.06    & non-type I PNe            & 72  \\
O/H               & -0.039$\pm$0.033 &  8.56$\pm$0.15    & type I PNe                & 19        \\ 
\hline 
Ne/H              & -0.037$\pm$0.018 &  7.75$\pm$0.07    & All PNe                   & 55              \\
Ne/H              & -0.051$\pm$0.019 &  7.76$\pm$0.08    & non-type I PNe            & 37   \\
Ne/H              & -0.025$\pm$0.037 &  7.76$\pm$0.17    & type I PNe                & 18        \\ 
\hline 
S/H              & -0.033$\pm$0.019 &  6.77$\pm$0.08    & All PNe                     & 38            \\
S/H              & -0.027$\pm$0.020 &  6.67$\pm$0.09    & non-type I PNe    & 19   \\
S/H              & -0.041$\pm$0.031 &  6.87$\pm$0.14    & type I PNe        & 19        \\ 
\enddata
\label{Tab_grad}
\tablecomments{(1) The chemical element for which the gradient is computed;  (2) slope of the abundance gradient;   
(3) zero-point (chemical abundances in the centre of the galaxy); (4) sample of PNe; 
(5) number of PNe in each sample. 
For each element, the first row report the gradient obtained
considering the whole sample of disk PNe (thus excluding the two
possible halo PNe), the second row gives the gradient computed using non-Type I PNe, 
and the third row gives the gradient obtained of  Type I PNe.}
\end{deluxetable}

\section{Discussion}            
\label{sect_discu}

The metallicity gradient of PNe allows to analyze how was the metal
distribution in the past epochs of M33. In M33 we have to possibility
to compare the metallicity gradient of PNe with a good number of other
measurements of abundance gradients, in particular from \hii\ regions.
The advantage of comparing PNe to \hii\ regions is the similarity of
their emission-line spectra in spite of their different evolutionary
states that allows to use the same observation techniques, analysis,
and abundance determinations.

\begin{figure}
\resizebox{\hsize}{!}{\includegraphics[angle=0,scale=1.00]{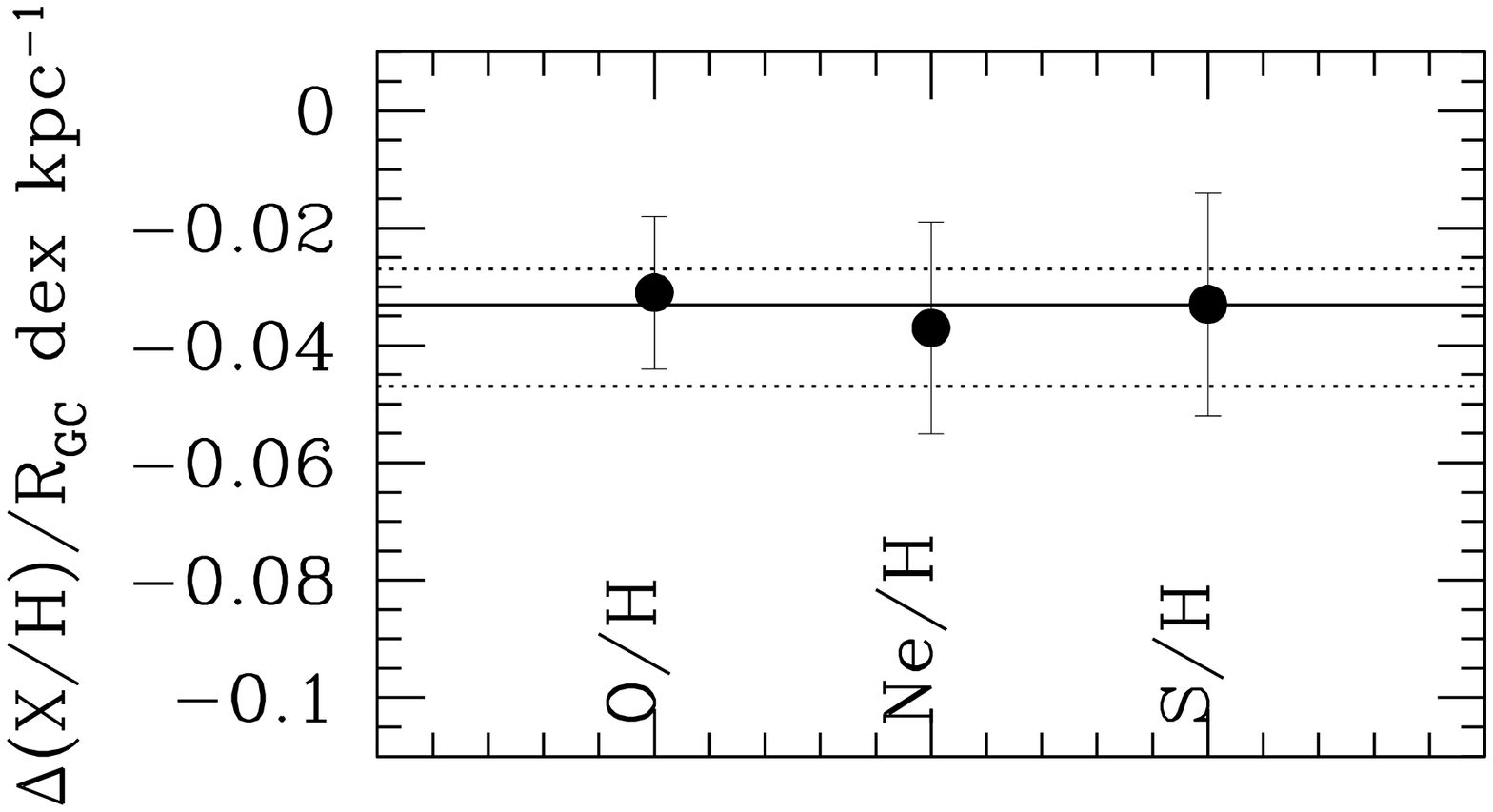}}
\caption{The slopes of the radial elemental gradients in M33. 
The continuous line is the weighted average slope, and the two dashed line 
give the rms $\sigma$ of this value. }
\label{Fig_grad}
\end{figure}

In order to compare our PN results with the leading samples of M33
\hii\ regions we build the largest, most reliable sample available to
date.  As already noted by e.g. RS08 and M07a, the size and quality of
the sample is fundamental to derive reliable metallicity gradients. In
\hii\ regions there is a substantial intrinsic scatter of 0.11 dex in
the metallicity at any given distance from the M33 center, which
imposes a fundamental limit on the accuracy of gradient measurements
that rely on small samples of objects (RS08).  Also, in the past, only
giant \hii\ regions have been studied, and those might have a steeper
metallicity gradient than small and compact \hii\ regions.

We have re-calculated the
\hii\ regions metallicity gradients considering a cumulative sample,
including that by RS08, consisting of 61 \hii\ regions with accurate 
abundance determinations, and that by M07a that included new and literature
spectroscopy performed to a set of 28  
\hii\ regions.
 The cumulative sample of 89 \hii\ regions gives,

$$ 12 + {\rm log(O/H)} = -0.032 (\pm 0.009) \times  {\rm R_{GC}} + 8.41 (\pm 0.04).\eqno(4)$$

When compared to the best gradient determined using M33 PNe (Eq. 3), we
confirm that these gradients are consistent within the errors.  We
infer that the metallicity gradient of M33 has not changed from the
epoch of the formation of the PN progenitors to the present time.

We can use for comparison the gradient of the best measured element, oxygen,
or the  
weighed average of the slope of O/H, Ne/H, and S/H gradient, and shown in
Fig.~\ref{Fig_grad},  
-0.033$\pm$0.003~dex kpc$^{-1}$.

In Table~\ref{Tab_grad_lit} we compare our metallicity gradient with
the literature values derived using different stellar populations.
$\alpha$-element and iron-peak gradients are separated by a double
horizontal line since their different origin (SNII vs. SNIa)
suspects a different metallicity distribution through the disk and
possibly a different gradient.  In each group, $\alpha$-elements or
Fe/H, gradients are ordered by the increasing age of the corresponding
stellar population. We considered only metallicity gradients computed
with a statically significant sample, thus, e.g., excluding the
metallicity gradient with Cepheid stars (Beaulieu et al.~2006).

It is remarkable that both the oxygen and neon metallicity gradients
of M33 do not change within the errors even considering populations of
different ages, though they are slightly different, the neon gradient
having a slightly steeper slope.  

In the following we discuss which evolutionary scenario and chemical
evolution models are excluded or favored by our results.

The first chemical evolution model attempting to the describe the evolution of M33 
was that by  Diaz \& Tosi~\cite{diaz84}. Their model assumes a constant and uniform infall
rate and fails to reproduce the metallicity gradient observed in M33.
Other models that describe the origin and evolution of metallicity gradients
are listed in Table 7, where we 
report the Reference (column 1), the model type (column 2), the resulting oxygen
gradient (column 3),  and its variation with time (column 4), where a plus
sign indicates a flattening of the  metallicity gradient with time.

The first three models listed in Table 7 are a family of multiphase models,
differing 
mainly from the origin and rate of the infalling material.
A multiphase model considers the galaxy divided in two or more zones,
typically the halo and the disk, 
composed by baryonic material in different phases: diffuse gas, clouds,
stars,  and stellar remnants. These models 
evolve with time (see e.g. Ferrini et al. 1992, 1994 for a complete
description), and 
are characterized by a flattening of the radial metal distribution with time,
due to the inside-out formation of the disk. 

Moll\'a et al.~\cite{molla97}  specifically analyzed the time-evolution
of the metallicity gradient in several spiral galaxies, where the disk is formed
by the primordial gas of the spheroidal protogalaxy which collapses onto the
galactic plane and forms out the disk. For 
M33 they obtained that $\Delta$(O/H)/$\Delta$(R)$\sim$ -0.21 dex kpc$^{-1}$,
flattening  with time at a rate of  +0.005 dex kpc$^{-1}$ Gyr$^{-1}$. 

Moll\'a \& Diaz~\cite{molla05} applied their results to M33 and found
metallicity gradients similar to those  
by Moll\'a et al.~\cite{molla97}. Gradient evolution is not computed
explicitly,  but a clear flattening of the gradient 
is evident in their figures. 

M07b built the so-called "accretion" model reproducing  the slow formation of
the disk of M33 from the inflow of intergalactic gas, 
which  predicts  $\Delta$(O/H)/$\Delta$(R)$\sim$ -0.067  dex kpc$^{-1}$, also
flattening  with time at a rate of  +0.003 dex kpc$^{-1}$ Gyr$^{-1}$. 

Other types of models, such as those by Chiappini et al. (1997, 2001), were
not applied directly to M33, but are notable
since they predict  steepening of the metallicity gradients with time. These
models 
assume two main accretion episodes for the formation of the galaxy, the first
forming the halo, the  bulge, and the thick disk  
in a short timescale, and the second forming the thin disk, with a timescale
that is an increasing function of the galactocentric distance. 
Different assumptions on the star formation threshold and on the timescales
of the infall episodes produce the gradients slopes and evolutions 
reported in  Table \ref{tab_model} (see Chiappini et al. 2001 for a detailed
description). 
Note that the behavior of Model D is different from the other models by the
same authors in that the assumed  slower formation of the outer halo   
does not influence the disk formation, resulting in steeper gradients.

If for example we assume that the age of all PNe is  5 Gyr, then according to
the models by Molla et al. we would  
expect a PN gradient which is -0.025 dex kpc$^{-1}$ steeper than that  of
the \hii\ regions. Using the M07b models for the same PNe age, the PN gradient
is -0.015 dex 
kpc$^{-1}$ steeper than that of the \hii\ regions. 
On the other hand, with Models A, B, C by Chiappini et al. (2001) we would
observe PN gradients that are   
flatter that those of \hii\ regions by +0.04, +0.02, +0.03 dex kpc$^{-1}$
respectively. 
These  variations could be even larger for older PN populations.
In summary, only the models by M07b, and
model B by Chiappini et al. \cite{chiappini01}, are marginally consistent
with the observations. 

To match the small metallicity gradient variation with time we observed with
the models of M33 we could invoke a lower accretion rate from the ISM than
that used in the M07b models. 
The resulting simulation should produce
a flat metallicity gradient, and a small slope variation in the last 8-10
Gyr, as observed. Detailed analysis of galactic chemical evolution models are
out of the scope of this work.

\begin{deluxetable}{cccc}
\tabletypesize{\scriptsize}
%\rotate
\tablecaption{Chemical evolution models of M33}
\tablewidth{0pt}
\tablehead{
\colhead{Ref.} & \colhead{Model} & \colhead{$\Delta$(O/H)/$\Delta$(R) (present time)} & \colhead{($\Delta$(O/H)/$\Delta$(R))/$\Delta$(t)} \\
                        &                            & (dex kpc$^{-1}$)                                                           & (dex kpc$^{-1}$ Gyr$^{-1}$)\\
(1) & (2) & (3) & (4)
}
\startdata
Moll\'a et al. (1997)      & multiphase, infall from collapse of the halo  & -0.21      & +0.005             \\
Moll\'a \& Diaz (2005)  & multiphase, infall from collapse of the halo  &-0.20       & ...                                                 \\
M07b                              & multiphase, accretion model                           &  -0.067   & +0.003          \\
\hline
  Chiappini et al.(2001) & model A   & -0.065  & -0.008      \\
                                         & model B    &  -0.045  & -0.004       \\
                                         & model C    & -0.070   & -0.006               \\
                                         & model D    &   -0.20    & +0.05      \\

\enddata
\label{tab_model}
\tablecomments{(1) Reference; (2) main characteristics of the chemical evolution model; (3) slope of the 
present-time O/H radial gradient;  (4) time variation of the O/H gradient. }
\end{deluxetable}

\begin{deluxetable}{cccccc}
\tabletypesize{\scriptsize}
%\rotate
\tablecaption{A sample of literature abundance gradients}
\tablewidth{0pt}
\tablehead{
\colhead{Element} & \colhead{Gradient} & \colhead{Population} & \colhead{Age} & \colhead{\#} & \colhead{Ref.}\\
(1)          & (2) & (3) & (4) & (5) & (6)
}
\startdata
O/H               & -0.05$\pm$0.025   & B giants              & few Myr                  &30 & Urbaneja et al.~\cite{urbaneja05} \\
O/H               & -0.032$\pm$0.009  & \hii\ regions         & few Myr                  &89 & RS08 + M07a \\
%O/H              & -0.08$\pm$0.06    & B giant               & few Myr                  &   & Monteverde et al. (2000) \\
O/H               & -0.039$\pm$0.033  & young PNe               & $<$0.3 Gyr                 &19& this paper                          \\
O/H               & -0.031$\pm$0.013  & All PNe               & $<$10 Gyr                 &91 & this paper                          \\
O/H               & -0.030$\pm$0.013  & old PNe               & 0.3-10 Gyr                 &72 & this paper                          \\
\hline
Ne/H              & -0.05$\pm$0.02    & \hii\ regions         & few Myr                  &30 & Willner \& Nelson-Patel~\cite{willner02} \\
Ne/H              & -0.058$\pm$0.014  & \hii\ regions         & few Myr                  &25 & Rubin et al.~\cite{rubin08} \\
Ne/H               & -0.025$\pm$0.037  & young PNe               & $<$0.3 Gyr                 &18& this paper                          \\
Ne/H              & -0.037$\pm$0.018  & All PNe               & $<$10 Gyr                 &55 & this paper  \\  
Ne/H              & -0.051$\pm$0.019  & old PNe               & 0.3-10 Gyr                 &37 & this paper  \\  
\hline 
\hline
Fe/H              & -0.06              & AGB                  & $\sim$4-6 Gyr           &- & Cioni et al.~(2008) \\
Fe/H              & -0.04$\pm$0.02     & RGB                  & $>$8 Gyr                &- & Kim et al.~(2002)\\
Fe/H              & -0.06$\pm$0.01     & RGB                  & $>$8 Gyr                &- & Tiede et al.~(2004)\\
\enddata
\label{Tab_grad_lit}
\tablecomments{(1) Chemical element; (2) slope of the radial gradient with its error; (3) stellar population; (4) typical age of 
the stellar population; (5) size of the sample; (6) reference.}
\end{deluxetable}

\section{Summary and conclusions}
\label{sect_conclu}

We present spectroscopic observations of a large sample
(102) of PNe in the spiral galaxy M33.  The observations were secured
with the spectrograph Hectospec on the 6.5~m telescope MMT.

For 32 PNe the electron temperature was directly computed. Empirical
relationships among electron temperatures and bright emission lines as
\heii\ $\lambda$4686, and \oii\ $\lambda$3727 were used to derive
electron temperatures of the remaining PNe.

Abundance diagnostic diagrams were built to study the PN population.
From the plot of N/O versus He/H we found that the PNe that fulfill the
Dopita~\cite{dopita91} definition of Type I PN are 19. 
Most PNe are non Type I, implying a population mainly composed by PNe
from old progenitors, with $<$3M${_\odot}$ and ages $>$0.3 Gyr.

A tight relationship between the O/H and Ne/H abundances was found, 
excluding the modification of both elements by PNe progenitors and ensuring
oxygen to be a good tracer of the galaxy metallicity.

The average chemical abundances of the PNe were compared to the PN population
of the LMC, the SMC, and the Galaxy. Generally speaking, the elemental
abundances of the $\alpha$-elements of the PNe in the disk of M33 are very
similar to those of the LMC.  The comparison between M33 PNe and \hii\
regions indicates a negligible global enrichment of the M33 disk
from the epoch of the formation of the PN progenitors to the present
time.

The radial metallicity gradients of those elements which are not
modified during the lifetime of LIMS-- oxygen, neon, sulphur-- were
derived.  The best measured element, oxygen, has a slope -0.031 ($\pm$
0.013) dex kpc$^{-1}$ in agreement within the errors with the same
gradient derived from \hii\ regions
-0.032 ($\pm$ 0.009) dex kpc$^{-1}$.

Since the metallicity gradients of PNe and \hii\
regions are practically indiscernible and the mean abundances of the PN
and \hii\ region populations are very close, we conclude that the chemical enrichment
in M33 from the time of the formation of the PN progenitors to the
present-time has been nearly negligible.

\appendix
\section{PN039: a planetary nebula with a [WC]-nucleus}

PN039 has a WR central star. Its spectrum shows prominent and wide
features of He~II at $\lambda$ 4686, 4541, 5411 and of He~I at
$\lambda$ 4471, 5876, 6678, 7065. In addition other stellar lines of
[CIII] at $\lambda$ 4056 and 4649, and [CII] at 7107 and 7124 were
detected.  The FWHM of these lines is 15\AA, to be compared
instrumental resolution of about 6 \AA.  Thus, following the
classification of Acker \& Neiner~\cite{acker03} PN039 has a late-type [WC]-9
or [WC]-10 nucleus due to the presence of very bright [CIII] and [CII]
lines respect to the [CIV] lines.

The ratio of the sulphur doublet indicates a low density limit
$\lesssim$100 cm$^{-3}$, and
\oiii,  4363/5007 ratio is completely anomalous for a normal PN, giving an 
electron temperature $\sim$ 30,000 K.
The reddening of the nebula is null, within the errors, indicating a little
presence of dust.

\begin{figure}
\resizebox{\hsize}{!}{\includegraphics[angle=0,scale=0.5]{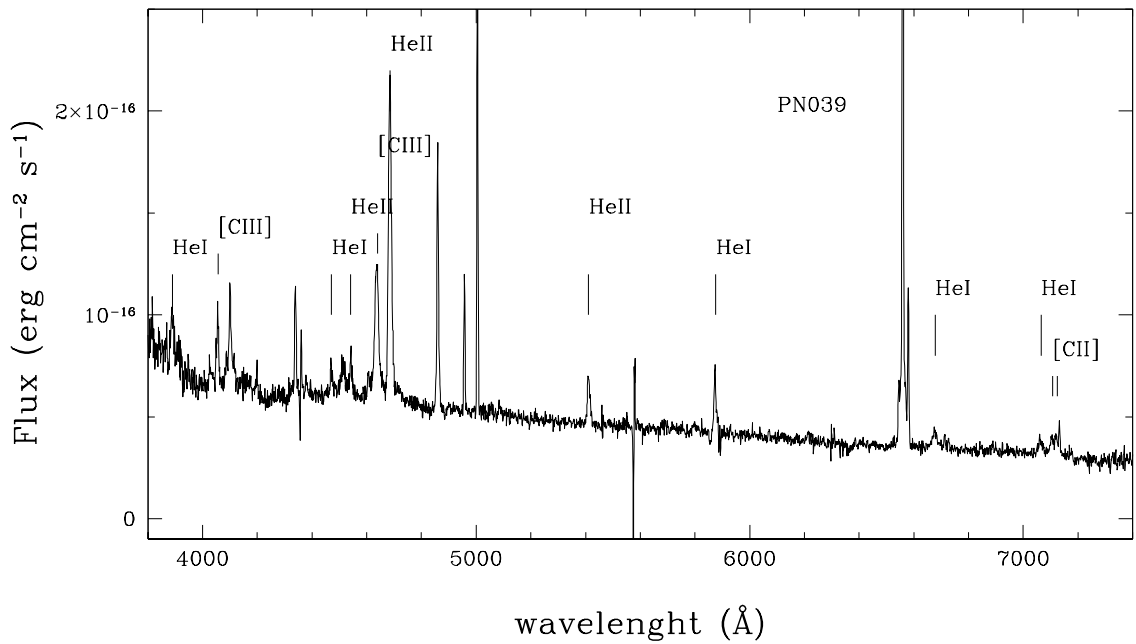}}
\caption{The spectrum of PN039 with its [WC] features. }
\label{Fig_pn39}
\end{figure}

\acknowledgments
{\em Acknowledgments:} 
We thank an anonymous referee for his/her valuable comments and suggestions that have improved the paper. 
We thank D. Fabricant  for making 
Hectospec  available to the community and  the TAC for awarding us the observing time.  
We thank the Hectospec instrument team and MMT staff for their expert help in
preparing and carrying out the Hectospec observing runs. 
We thank N. Caldwell, D. Ming and their team for the help during the data reduction. 
Thanks to Katia Cunha for enlightening scientific discussion, and Richard Shaw for his help with
the {\it nebular} package.

{\it Facilities:} \facility{Hectospec (MMT)}.

\clearpage

\end{document}